\documentclass[11pt]{article} 
\usepackage{hyperref}  
\pdfoutput=1 
\begin{document} 
\title{Dropping the Ball:\\
        The effect of anisotropic granular materials on ejecta and impact crater shape\\
	Entry\#: 102236} 
\author{Philip Drexler, Nathan Keim$^\dagger$ \& Paulo Arratia$^\dagger$ \\
\\
Department of Physics \\ 
Haverford College, Haverford, PA, 19041 USA \\
$^\dagger$ Department of Mechanical Engineering and Applied Mechanics\\
University of Pennsylvania, Philadelphia, PA, 19104 USA} 
\maketitle 
\begin{abstract} 
In this fluid dynamics video, we present an experimental investigation of the shape of impact craters in granular materials. Complex crater shapes, including polygons, have been observed in many terrestrial planets as well as moons and asteroids \cite{Ohman2010}. We release spherical projectiles from different heights above a granular bed (sand). The experiments demonstrate two different techniques to create non-circular impact craters, which we measure by digitizing the final crater topography.  In the first method, we create trenches in the sand to mimic fault lines or valleys on a planetary target. During impact, ejecta move faster in the direction of the trenches, creating nearly elliptical craters with the major axis running parallel to the trench.  Larger trenches lead to more oblong craters.  In the second method, a hose beneath the surface of the sand injects nitrogen gas.  The pressure of the gas counters the hydrostatic pressure of the sand, greatly reducing static friction between grains above the injection point, without disturbing the surface.  The affected sand has lower resistance to impact, creating a knob in the otherwise circular crater rim.
\end{abstract} 
%

\end{document}